\documentclass[final]{IEEEtran}

\usepackage{amsthm,amssymb,graphicx,multirow,amsmath,color,amsfonts}
\usepackage{subfigure}
\usepackage[update,prepend]{epstopdf}
\usepackage[noadjust]{cite}
\usepackage[latin1]{inputenc}
\usepackage{tikz}
\usetikzlibrary{arrows,calc}		% Optional ticks libraries
\usepackage{bbm} % for \mathbbm{1}
\usepackage{pdfpages}
\usepackage{booktabs}
\usepackage{tabulary}
\usepackage{multirow}
\usepackage{comment}
\usepackage{textcomp}
\usepackage[bottom]{footmisc}
\usepackage{hyperref}
\usepackage{enumitem}
\newcommand\footnoteref[1]{\protected@xdef\@thefnmark{\ref{#1}}\@footnotemark}

\include{notation}
\allowdisplaybreaks % Allows breaking of eqnarray over multiple pages (avoids unnecessary blanks in the document before eqnarray)

\renewenvironment{IEEEbiography}[1]
  {\IEEEbiographynophoto{#1}}
  {\endIEEEbiographynophoto}

\begin{document}
\title{Predictive Closed-Loop Service Automation in O-RAN based Network Slicing}
\author{
Joseph Thaliath, Solmaz Niknam, Sukhdeep Singh, Rahul Banerji, Navrati Saxena, Harpreet S. Dhillon, Jeffrey H. Reed, Ali Kashif Bashir, Avinash Bhat and Abhishek Roy
\thanks{J. Thaliath, S. Singh, R. Banerji and A. Bhat are with Samsung R\&D, Bangalore, India ({\texttt {email: \{jo.thaliath, sukh.sandhu, r.banerji, avinash.bhat\}@samsung.com}}. S. Niknam, H. S. Dhillon, and J. H. Reed are with Virginia Tech, Blacksburg, VA ({\texttt {email: \{slmzniknam, hdhillon, reedjh\}@vt.edu}}).
N. Saxena is with San Jose State University, CA, USA ({\texttt {email: navrati.saxena@sjsu.edu}}).
A.K. Bashir is with Manchester Metropolitan University, U.K. ({\texttt {email: a.bashir@mmu.ac.uk}}).).
A. Roy is with MediaTek, San Jose, CA ({\texttt {email: abhishek.roy@mediatek.com}}).).
%S. Yoon is with Samsung Electronics, Suwon, South Korea ({\texttt {email: siyoon72@samsung.com}}). \newline
This work is supported in part by the U.S. NSF (Grants CNS-1564148,  CNS-1814477), Oak Ridge National Laboratory (Grant 4000170832), and Commonwealth Cyber Initiative.}}
\maketitle

\begin{abstract}
Network slicing provides introduces customized and agile network deployment for managing different service types for various verticals under the same infrastructure. To cater to the dynamic service requirements of these verticals and meet the required quality-of-service (QoS) mentioned in the service-level agreement (SLA), network slices need to be isolated through dedicated elements and resources. Additionally, allocated resources to these slices need to be continuously monitored and intelligently managed. This enables immediate detection and correction of any SLA violation to support automated service assurance in a closed-loop fashion. By reducing human intervention, intelligent and closed-loop resource management reduces the cost of offering flexible services. Resource management in a network shared among verticals (potentially administered by different providers), would be further facilitated through open and standardized interfaces. Open radio access network (O-RAN) is perhaps the most promising RAN architecture that inherits all the aforementioned features, namely intelligence, open and standard interfaces, and closed control loop. Inspired by this, in this article we provide a closed-loop and intelligent resource provisioning scheme for O-RAN slicing to prevent SLA violations. In order to maintain realism, a real-world dataset of a large operator is used to train a learning solution for optimizing resource utilization in the proposed closed-loop service automation process. Moreover, the deployment architecture and the corresponding flow that are cognizant of the O-RAN requirements are also discussed.
\end{abstract}

%\begin{IEEEkeywords}
%O-RAN, 5G, RAN slicing, machine %learning, closed-loop automation, %resource provisioning
%\end{IEEEkeywords}
%%%%%%%%%%%%%%%%%%%%%
\section{Introduction} \label{sec:intro}
%%%%%%%%%%%%%%%%%%%%%
Compared to the rigid legacy wireless architectures, fifth generation (5G) technology promises to provide customized networks for vertical industries (like automotive, healthcare, agriculture, city management, manufacturing, etc.) with diverse service requirements. These diverse vertical services bring about a wide range of performance requirements in throughput, capacity, latency, mobility, reliability, position accuracy, etc. 3GPP Rel-15 standards~\cite{3GPP:TS22.261} introduced the concept of Network slicing, for providing the foundation of a common connectivity platform for these different services and satisfying the diverse requirements of these services. Network slices are logical networks that are provisioned with a set of isolated virtual resources on top of a shared physical infrastructure ~\cite{robust2019wen}. Network slicing can be realized by creating logical network functions and decoupling them from physical infrastructure. Currently, 3GPP Rel-17 standardization~\cite{3GPP:TS22.261} activities on RAN slicing envision to empower 5G network slicing as a tool that operators can use to open up new source of revenue by involving the application providers in customization of 5G wireless design, deployment and operation for better support of the applications.

With resources being deployed as virtual machines rather than physical hardware, automation of service assessment and resource orchestration are important. Automation enables immediate recognition and correction of undesired events over network slices, without explicit human intervention. This evolution of zero-touch service assurance and closing the loop between recognition and correction significantly reduces the cost of creating flexible services adapted to the needs of customers~\cite{Auto2019kafle}. However, providing such agility in the slice deployment, along with heterogeneity in topology, creates challenges in resource orchestration and dynamic service configurations~\cite{chochliouros2020dynamic}. Hence, advanced techniques are required to achieve optimal resource utilization while satisfying the requirements of multitude of services.

Given the development of machine learning (ML) techniques, intelligence is becoming a viable option for handling the heterogeneity of 5G wireless, including network slices~\cite{smart2019Xing}. The telecom industry has already demonstrated the use of ML in many trial deployments. Interestingly, 3GPP standardization has recently decided to look at ML with two different lenses: (a) one for the network and (b) one for the radio interface. The new 3GPP Rel-17 Study Item (SI) on the applications of ML to 5G RAN is just completed, and Rel-18 Work Item (WI) for standardization is approved to start from 2022~\cite{3GPP:TS37.817}. A separate 3GPP Rel-18 SI on application of AI-ML for improving the efficiency of wireless channels and radio interfaces is also approved to start from 2022~\cite{RP213469}. Hence, the de-facto wireless standardization is already poised to embrace ML for future 5G/B5G wireless. On the other hand, the Open Radio Access Network (O-RAN), an emerging new Radio Access Network (RAN) architecture, %\footnote{O-RAN is proposing white box concept wherein different service providers are able to inter-operate. With O-RAN gaining popularity amongst the operators, which is evident from more than hundred operators joining the O-RAN forum, service providers have to make their products compliant to O-RAN specifications. White box concept of O-RAN makes it easier for the operators to stitch modules coming from different service providers.}
which builds on two core pillars that are intelligence and open interfaces, introduces specific ML-based functions like the Radio Intelligent Controller (RIC)~\cite{niknam2020intelligent}.
%In that case, there is a strong need to identify existing standardization gaps and architectural limitations that are essential for a large-scale adoption and deployment of the new architecture. Hence, the evaluation of this new architecture, including the RIC to perform the optimization of features like Network slicing, is a must.

In this article we provide an ML-based resource-provisioning scheme for Network slicing in the O-RAN framework to prevent Service Level Agreement (SLA) violations. A concrete practical example is discussed to explain details on the deployment architecture and the end-to-end workflow. In this intelligent scheme, AI techniques are used to perform predictions of future SLA violations and perform corrective actions in advance. Specifically, a recurrent neural network model is utilized to predict the amount of resources required over each slice, given the volume of traffic it carries. End to end O-RAN setup has been used for evaluation of the intelligent closed control loop, resource provisioning scheme, for Network slicing and control the physical resources and the cloud resource of the slices, respectively. This closed control loop operation continues until the defined SLAs are met. It is worth clarifying that the proposed O-RAN based intelligent and automated setup is the first proposal of it's kind that can help in dynamically optimizing the over-served and under-served users in a slice, as the current implemented approaches in real-world networks are static solutions.

%%%%%%%%%%%%%%%%%%%%%
\section{Preliminaries}  \label{sec:Prelim}
%%%%%%%%%%%%%%%%%%%%%
5G marks the transition to a more dynamic network architecture by introducing the concept of \textit{slicing}. While the exact implementation of network slicing is operator specific, standards development organizations (SDOs) have provided template frameworks to facilitate this implementation~\footnote{ORAN Alliance, ``O-RAN working group 1: Study on O-RAN slicing'', Tech. Rep., Oct. 2020.}. A general framework of slicing implementation consists of two main phases i) slice instantiation and ii) resource orchestration over slices. During the first phase, a suitable set of network functions are configured to create a certain slice based on the defined slice specifications. Subsequently, the desired infrastructure resources for these network functions are allocated to meet the slice SLA requirements~\cite{Shen2020AIslice}. The isolation of resources on each slice is the key in preventing SLA violation between the slices. In general, there are two approaches for slice isolation: static and dynamic. In the static slicing, full isolation of the resource is guaranteed, {\em albeit} at the cost of possible over-provisioning of the resources. On the other hand, in a dynamic scenario, resources are multiplexed among the slices, by using advanced resource sharing and provisioning techniques.

\begin{figure}[b]
%\captionsetup{justification=centering}
\centerline{\includegraphics[scale=0.38]{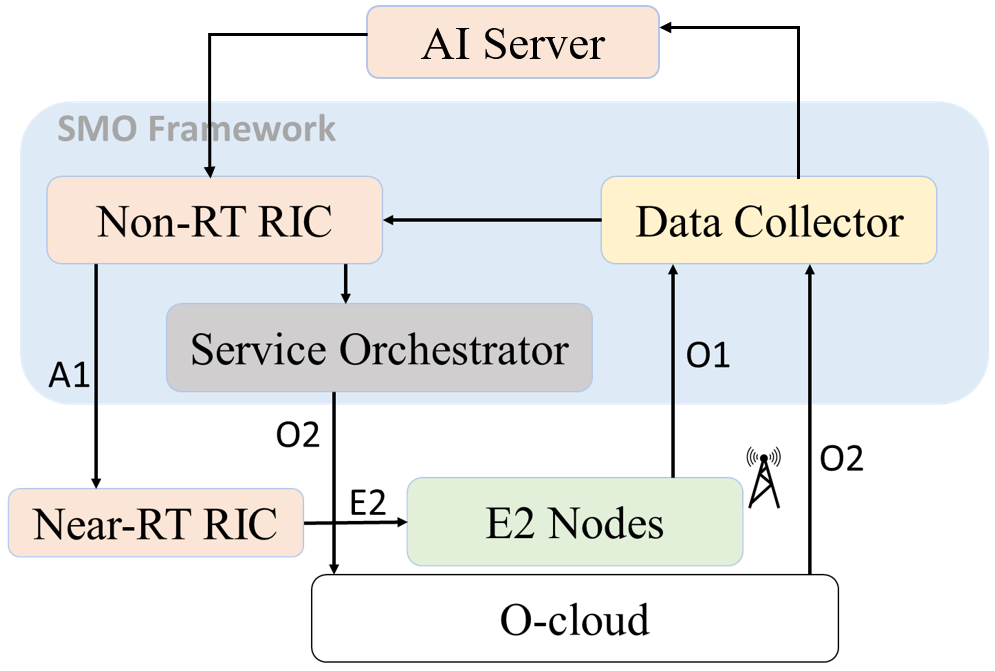}}
\caption{Intelligent and closed-loop resource provisioning for O-RAN slice automation.}
\label{fig:closedLoop}
\centering
\end{figure}

%\begin{figure*}[t]
%%\captionsetup{justification=centering}
%\centerline{\includegraphics[scale=0.6]{Flow.pdf}}
%\caption{Flow of the proposed intelligent automated resource orchestration over slices for O-RAN architecture.}
%\label{fig:Flow}
%\centering
%\end{figure*}
As networks are becoming more dynamic and the resources are more heterogeneous, ML is needed to handle optimization of the networks and resources. It enables automation of resource modification and adaptation without human intervention. An automated network is based on a closed-loop control with adaptability enabled through feedback~\cite{van2018automate}. Many initiatives are focused on automated resource management for network slicing by using open platforms and open source tools, with the objective to provide automation across potentially multi-tenant slices and simplify resource orchestration~\cite{boub2019lifecycle}. Given that the O-RAN architecture is based on the principles of \textit{openness} and \textit{intelligence}, it can assist in providing predictive and closed-loop service automation by intelligently and dynamically orchestrating resources over the slices and maintain resource isolation.
O-RAN architecture aims to reduce the proprietary implementation of hardware and software which in turn helps in increasing operational savings by establishing open standards. Additionally, openness of RAN components expedites the provisioning of new services for users. Moreover, intelligent RAN also reduces human intervention in the loop and increases the accuracy to handle the network complexity. The following are the main functional entities and interfaces (Fig. ~\ref{fig:closedLoop}):
\begin{enumerate}
\item Non-RT RIC: The Non-Real Time (Non-RT) RIC (Radio Intelligent Controller), which is part of the Service and Management Orchestration (SMO) system, provides a platform to deploy optimization applications, called rApps.
\item Near-RT RIC: The Near-Real Time (Near-RT) RIC, which is deployed closer to RAN nodes, provides another platform to deploy optimization applications, called xApps.
\item E2 Nodes: These nodes correspond to infield deployed RAN nodes supporting O-RAN standardized interfaces, like E2 and O1. E2 interface is between Near-RT RIC and E2 Nodes and is used by Near-RT RIC to configure E2 nodes based on policy formed at Near-RT RIC. O1 interface is between E2 nodes and the data collector. It is generally used by Service Management and Orchestration (SMO) Framework to collect RAN Performance Management (PM) statistics from E2 nodes.
\item O-Cloud: O-Cloud corresponds to cloud computing platform where some of the functions of the RAN nodes and Near-RT RIC are deployed. There is an O2 interface between O-Cloud and SMO Framework. It is utilized to collect cloud statistics through data collector in SMO. It is also utilized to configure O-Cloud with the help of Service Orchestrator (SO) through policies formed at the Non-RT RIC.
\item AI Server: AI Server hosts the platform for training machine learning models required for RAN optimization. For that purpose, we have utilized a dataset captured from E2 nodes of a real-world cellular network. The required data are collected through the data collector located in the service management and orchestration SMO layer. The intelligent prediction modelis trained in a proprietary AI server and deployed in the Non-RT RIC as rApp.
\end{enumerate}
%Given that the emerging O-RAN architecture is primarily driven by the two main principles of \textit{openness} and \textit{intelligence}, it would be a suitable choice to facilitate predictive and closed-loop service automation by intelligently and dynamically orchestrate resources over the slices to maintain an acceptable level of isolation. Briefly speaking, O-RAN architecture aims to reduce the proprietary implementation of hardware and software which in turn helps in increasing operational savings by establishing open standards.
The two main components, to embed intelligence in O-RAN architecture and enhance the functionality of the traditional networks, are radio intelligent controllers, namely non-RT RIC and near-RT RIC. For the purpose of slicing, non-RT RIC collects key performance indicators (KPIs) related to slices, as well as the parameters that are used to configure them. These parameters and performance metrics are also sent to near-RT RIC to apply dynamic slice optimization for achieving SLA assurance. Based on the decision, network resources are controlled by near-RT RIC through the E2 interface and cloud resources by non-RT RIC through O2 interface~\footnote{https://www.o-ran.org\label{note2}}.
Please refer to~\cite{niknam2020intelligent}, for detailed description of O-RAN goals, architecture and challenges.

\begin{figure*}[h]
%\captionsetup{justification=centering}
\centerline{\includegraphics[scale=0.65]{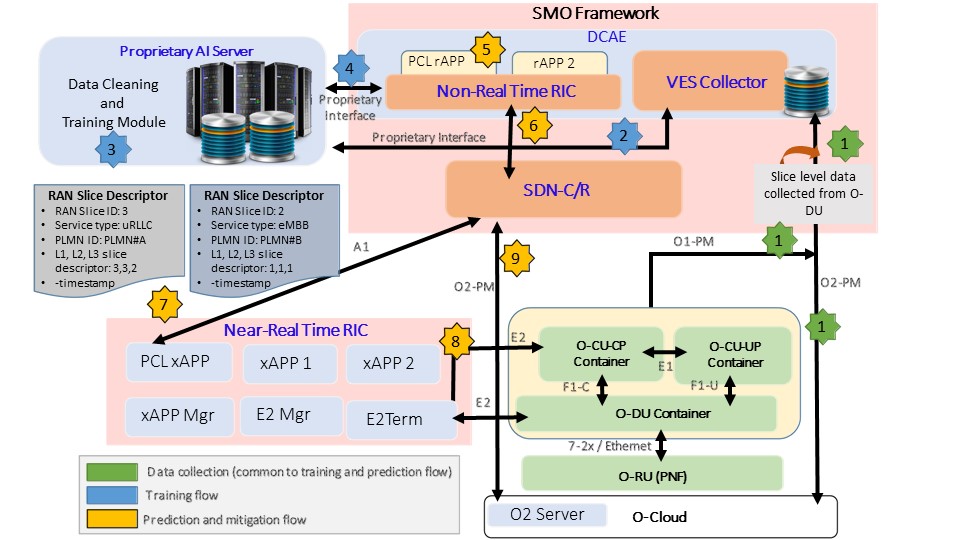}}
\caption{Structure of deploying the proposed predictive closed-loop service automation in O-RAN slicing.}
\label{fig:Arch}
\centering
\end{figure*}
%%%%%%%%%%%%%%%%%%%%%%%%%%%%%%%%%%%%%%%%%%%%%%%%%%%
\section{Deployment Architecture}
\label{sec:Proposed}
%%%%%%%%%%%%%%%%%%%%%%%%%%%%%%%%%
%O-RAN is in very nascent stage, where the architecture, interfaces and use cases are being discussed along with integration of network slicing concept. Moreover, the software community of O-RAN is strengthening it's basic component development. As discussed in Section II, dynamic resource provisioning in network slices is crucial to cope up with increasing heterogeneity in the 5G/B5G network.
With inspirations from software community we go a step ahead to integrate ORAN with real network components, along with our proprietary AI server, to form an inter-operable, end-to-end setup for testing dynamic resource provisioning in network slicing, using real world data. Our proposed RIC applications (i.e., Predictive Closed Loop (PCL) rAPP and PCL xAPP) checks the aforesaid network slicing use case in a closed loop automation environment. We believe this will further strengthen the O-RAN and network community to tackle stringent network slice resource management requirements, without the intervention of operators and prevent the SLA violation of the network slice(s). Our setup consists of wide range of Base Stations with variety of traffic patterns to make our solution more realistic and widely implementable. Fig.~\ref{fig:Arch} represents the structure of deploying the proposed predictive closed-loop service automation.
%This section provides the overall deployment and flow of proposed slicing scheme in the O-RAN architecture. It is worth noting that while providing architectural details, we also consider cloud resource optimization along with network resource optimization. The detailed description of the proposed learning scheme is provided as follows.

%We start by describing the structure of deployment scenarios and flow of the proposed slicing solution in the O-RAN architecture and also explain the implementation of proposed slicing in a closed-loop and automated fashion.   %Subsequently, the flow of the resource orchestration over slices is represented in Fig.~\ref{fig:Flow}. %The subsequent bullet points (marked by numbers on Fig.~\ref{fig:Arch}) can be followed in order.
Our real-world demo testbed includes open network automation platform (ONAP) and O-RAN software community projects hosted by Linux foundation that are publicly available to demonstrate the Predictive Closed Loop Automation. ONAP provides all the necessary functionalities related to service management and orchestration. The ONAP framework also provides an adapter to integrate proprietary AI server that is necessary to perform data training etc.
%The Frankfurt release of ONAP is considered for the setup\footnote{  https://wiki.onap.org/}.
This is expected to depict the Non-RT RIC functionality as listed in WG2 of O-RAN Alliance\footref{note2}. O-RAN SC provides all the necessary functionalities for Near-RT RIC, simulating the E2 Nodes (O-DU, O-CU etc.). The Cherry Release of the O-RAN SC is considered for the setup\footnote{ https://docs.o-ran-sc.org/en/latest}. The network slices comprise of these E2 nodes. (Note that: O-DU (Open-DU) is the distributed unit that sits close to the radio unit and runs the Radio Link Control (RLC), Medium Access Control) MAC, and parts of the Physical (PHY) layer and O-CU (Open-CU) is the centralized unit that runs the Radio Resource Control) RRC and Packet Data Convergence Protocol (PDCP) layers. The information flow between components inside ONAP will flow through Data movement as a platform ${(DMaaP)}^{2}$ which is a premier platform for high performing and cost effective data movement services that transports and processes data from any source to any target with the format, quality, security, and concurrency).
\begin{enumerate}
\item Considering the network slicing feature is not yet deployed by any operator in the world, we have mapped the existing 4G network data. Mapping of this real world data set to network slicing will be discussed in the next section. Field data collected from operator network is reported periodically via the O1 interface (defined in Section 2) to the Virtual Event Streaming ${(VES) \; collector}^{2}$ similar to that reported by an eNB in a commercial network data. Virtual Event Streaming (VES) Collector is a RESTful collector for processing JavaScript Object Notation (JSON) messages. (Please note that JSON is a text-based data interchange format to maintain the structure of the data). The related RAN statistics for each slice, including the number of active users per slice, volume of data and Physical Resource Blocks (PRB) utilization per slice are collecting in this manner. It is worth noting that there will be minimal processing for converting counters to KPIs and tag slice identifier (ID) at the data collector. Additionally the cloud resources related statistics are collected, such as virtual machines (VMs), computing units, memory units, from O-cloud (defined in Section 2) over O2 interface (defined in Section 2).
\item The KPI information is sent to the proprietary AI Server which is installed on a different server where data cleaning, data training and data prediction happens.
\item In order to learn the network or cloud usage pattern and predict the future values of the target parameters, an ML model is trained in the AI server. The trained model is utilized to predict the related KPIs which reflects the priority of the slice based on the network and cloud resource usage patterns.
\item The outcome of the prediction is sent back to predictive closed loop (PCL) rAPP (rAPP defined in Section 2) that is running as an application in Non-RIC platform.
\item PCL rAPP is an application host, part of Non-RT RIC. It also coordinates the formation of inference based on KPI prediction data and requirement of different slices based on slice level SLAs for network and cloud resources. PCL rAPP also prepares the RAN slice descriptor based on inference for configuring network resources. RAN slice descriptors contain slice IDs, public land mobile network (PLMN) ID, layer level individual descriptor and time stamp. The layer level network parameters that need to be scaled up/down in the individual layer slice descriptor and time stamps, can also be provided.
\item PCL rAPP forms the prediction results for cloud resources and shares it with SDN-C/R which is responsible for configuration O-Cloud. (Please note that SDN-C/R is a Software Defined Network controller which is built on the common controller framework that assigns, manages and provisions network resources with the help of radio interfaces.)
\item RAN slice descriptors will be passed on, over A1 interface by A1-controller, as a JSON policy to xAPP (xAPP defined in Section 2) using xAPP manager in Near-RT platform. The function of xAPPs applications is to provide programming capability for RAN component (O-CU/O-DU).
\item xAPP will trigger the scaling up/down of network resources as defined in RAN slice descriptor by sending control message over E2 with the help of E2 Term and E2 Manager (inside Near-RR RIC platform). We define events in a way to change their parameters in E2 termination function (inside Near-RT RIC platform). Event definition is based on in the ASN format. It is a control message that specifies the event and state. Based on the proposed scheme, we require to define the new values related to PRB utilization and number of users per slice based on the prediction results prepared (by rAPP) with the help of predicted KPIs and their respective SLAs.
\item On the other hand, O2 Client (inside SDN-C/R) helps in configuring (scale up/scale down) the cloud resources directly through O2 interface based on the inference output using O2 server in O-Cloud.  If, due to prediction in a particular time instance, the PCL rAPP decides that the particular service is of less priority, it can bring down the cloud resources of a slice via O2 interface with the help of SDN-C/R. The deactivated cloud resources can also be activated based on the requirement of the service using same interfaces.
\end{enumerate}

%\begin{figure*}
%     \centering
%     \begin{subfigure}
%         \centering
%         \includegraphics[scale=0.35]{Active_UE_LSTM.png}
%         \caption{\chr{Prediction results for Active number of UEs in one slice using LSTM approach.}}
%         \label{fig:LSTM}
%     \end{subfigure}
%     \begin{subfigure}
%         \centering
%         \includegraphics[scale=0.35]{Active_UE_ARIMA.png}
%         \caption{\chr{Prediction results for Active number of UEs in one slice using ARIMA approach.}}
%         \label{fig:ARIMA}
%     \end{subfigure}
%\end{figure*}
%%%%%%%%%%%%%%%%%%%%%
\section{Proposed Framework: Intelligent Resource Provisioning}
%%%%%%%%%%%%%%%%%%%%%
We now introduce our intelligent and automated resource provisioning scheme tailored for the O-RAN slicing architecture. Our proposed scheme learns temporal patterns of the traffic over four different traffic bearers with different service requirements by utilizing a deep learning based neural network. Such a learning is used to predict the number of users, volume of data and resource utilization for different cells in the network. To prevent a potential SLA violation, network resources are re-allocated accordingly. In fact, the goal is to utilize a commercial off-the-shelf (COTS) learning approach and incorporate intelligence in O-RAN architecture to proactively predict the resource requirements for efficient slicing.

Similar to legacy wireless networks, in 5G, standardized QoS identifier (5QI) are specified for bearers carrying different type of traffic corresponding to different types of services. The corresponding values of QoS to each identifier are provided in~\cite{3gpp.23.203}. Therefore, bearers with different 5QI can be considered as RAN slices, as they are carrying traffic of different service types. Therefore, bearers with different 5QI can be observed as a way of representing a slice given the limited implementation of slicing in real-world networks as of today.

\subsection{ML Model Architecture and Settings}
In this scheme, the number of active users for each bearer is predicted for the next prediction time window. Moreover, the actual and the required number of physical resource blocks (PRB) for each bearer are considered, as well. Based on the minimum amount of the required PRBs for accommodating the predicted number of active users on each bearer, resources are provisioned in anticipation of SLA violation.
In fact, to prevent QoS degradation, resources are scaled down/up among bearers based on the predicted amount of PRBs required for each bearer. It is worth noting that our primary goal is more focused on predicting the lack of balance of resources in  the  network, while providing  the  prediction result to manage the resources sufficiently ahead of time, by using RIC controllers. The set of subsequent approaches to mitigate the resource under/over-utilization is more related to the implementation of network slicing of RAN which is beyond the scope of this work.
In order to calculate the average utilization of PRB over each bearer, we utilize three other KPIs, namely active user per traffic bearer, total volume of data carried over each bearer and total downlink PRB (DL-PRB) utilization, which provides the percentage usage of downlink PRBs for user plane traffic~\cite{ETSIts132}. Moreover, the number of active UEs in the DL per 5QI is defined as the mean number of active DRBs for UEs in the DU. In addition, data volume per 5QI is the data volume (amount of PDCP SDU bits) in the downlink delivered to PDCP layer.
%It is notable that we target these other KPIs as the actual PRB utilization per bearer is not available in our dataset.

%%%%%%%%%%%%%%%%%%%%%%%%%%%%%%%%%%%%%%%%%%%%%%%
\begin{table}[t]
\caption{Simulation Parameters}
\label{table:sim_param}
\centering
\begin{tabular}{l|c}
\hline
\multicolumn{2}{c} {\textbf {LSTM Parameters}}  \\
\hline
\multicolumn{1}{c}{\textbf{Parameter}}  & \textbf{Value} \\
\hline
LSTM layers & 2 \\
Num. of LSTM units/layer &	150 \\
Activation fx &	relu \\
Batch Size & 24 \\
Num. of epochs	& 120 \\
Optimizer &	Adam \\
Accuracy Tolerance range	 &	1\% of max value (to avoid overfitting) \\
\hline
\multicolumn{2}{c} {\textbf {ARIMA Parameters}}  \\
\hline
\multicolumn{1}{c}{\textbf{Parameter}}  & \textbf{Value} \\
\hline
order ($p$,$d$,$q$,$P$,$D$,$Q$,$s$)	&	get from best fit model from autoarima \\
window\_size	&	24 \\
Accuracy Tolerance Range	&	1\% of max value (to avoid overfitting) \\
\hline
\multicolumn{2}{c} {\textbf {Server Details}}  \\
\hline
\multicolumn{1}{c} {\textbf{Parameter}}  & \textbf{Value} \\
\hline
Num. of eNB & 17 \\
Num. of cells in each eNB &	18 \\
Server & Dual Xenon Gold CPU \\
RAM & 512 GB \\
CPU	& 48 cores with 2 threads per core \\
RNN API used &	Keras \\

\hline
\end{tabular}\label{tab:sim_param}
\end{table}
%\vspace{0.5cm}

The simulation parameters are tabulated in Table~\ref{table:sim_param}, where $p$: trend autoregression order; $d$: trend integration order; $q$: trend moving average order; $P$: seasonal autoregression order; $D$: seasonal integration order; $Q$: seasonal moving average order; $s$: no. of time steps for a single seasonal period. The dataset, considered for the present simulation, belongs to a real cellular networks in India. From the total data set, $80\%$ of the data is considered as training set and $20\%$ is considered as test set. It contains network measurements in terms of number of active users for bearers, volume of data in GB, DL-PRB utilization percentage, collected from $17$ LTE eNBs ($18$ cells in each eNB), over a duration of $31$ days.  We consider $4$ different bearers including QCI1, QCI2, QCI5 and QCI9 in the dataset. Examples of services with these QoS identifiers can be, respectively, conversational voice, live video streaming, IP multimedia subsystem (IMS) signalling and buffered video streaming~\cite{3gpp.23.203}. The 5QI in 5G are similar to QCI in LTE~\cite{3gpp.23.501} The model training is performed and the traffic of the next hour is predicted where based on the periodicity of the available data, the prediction window can be configured accordingly. While the prediction is done on an hourly basis, but when the prediction process is done the decision time is in order of a few seconds. LSTM is an artificial recurrent neural network (RNN) architecture used in the field of deep learning. Unlike standard feedforward neural networks, LSTM has feedback connections. It can not only process single data points, but also entire sequences of data. In fact, the special structure of LSTM unit makes it more effective in capturing the dependency of neighboring data points in sequential data while avoiding vanishing/exploding gradient issues. As traffic data is a form of time sequence data in nature, LSTM has been a proper choice in dealing with traffic data. There are of course conventional approaches for dealing with sequential data such as autoregressive integrated moving average (ARIMA) and its variants, and we tested our data using this approach as well as LSTM. As a comparison, in LSTM approach we have training and testing accuracy of $91\%$ and $86\%$, respectively. In addition, the accuracy of ARIMA approach is $75\%$.
Based on the data-set we have, the LSTM approach works well in the prediction stage. It is worth mentioning that our goal is not to provide contributions and comparisons in the ML algorithm of choice, but rather to utilize a commercial off-the-shelf (COTS) learning approach and incorporate intelligence in O-RAN architecture to proactively predict the resource requirements for efficient slicing.

\begin{figure*}[t]
%\captionsetup{justification=centering}
\centerline{\includegraphics[scale=0.57]{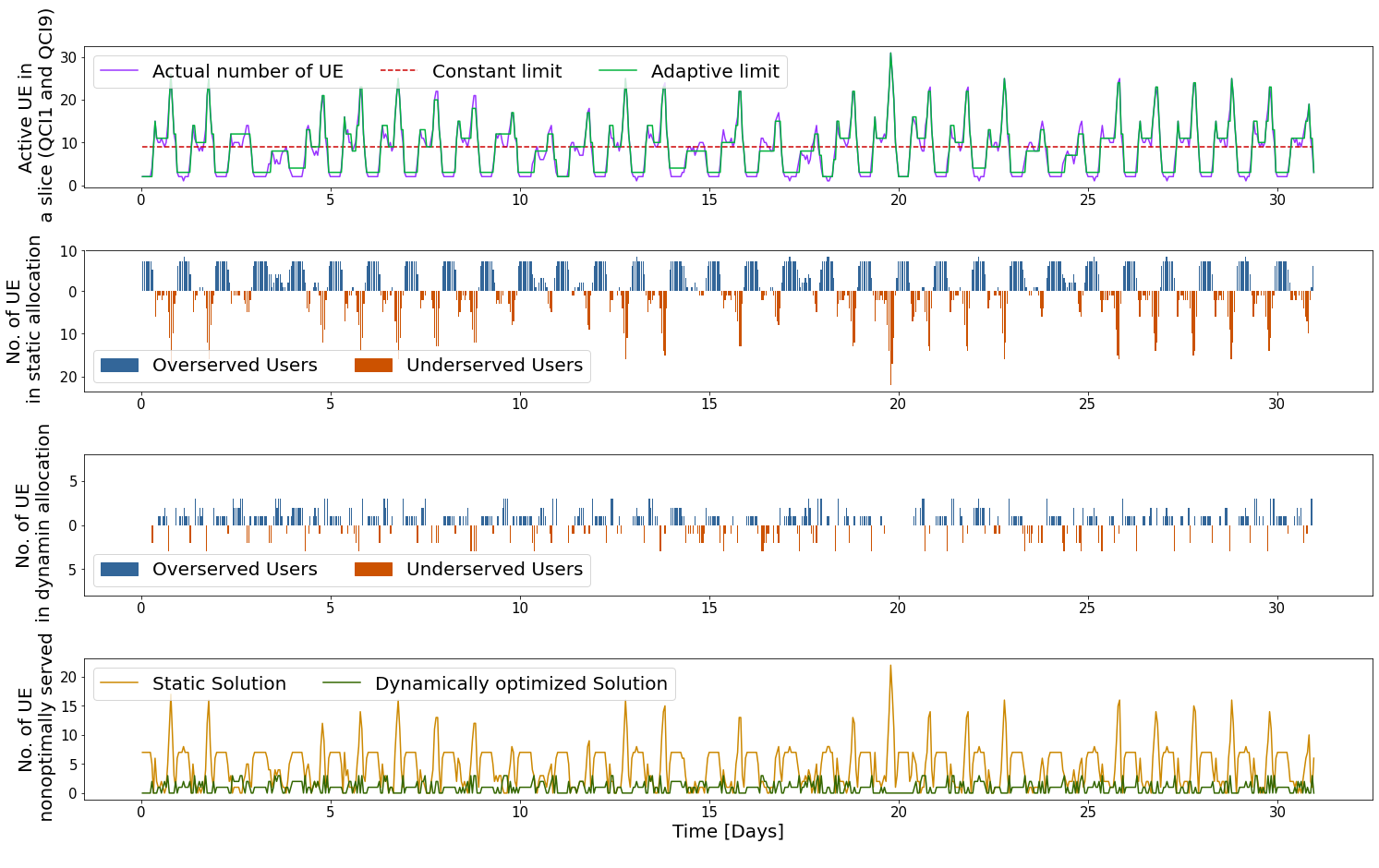}}
\caption{Performance of the proposed intelligent and automatic resource provisioning scheme over slices. The first subgraph illustrates number of active UEs over QCI1 (purple solid line), the constant (dotted red line) and adaptive (green line) limits on the number of UEs allowed in this slice. The second (resp. third) subgraph represents the number of under or over-served UEs in the static (resp. dynamic) resource provisioning scenario. Last subgraph represents the overall number of UEs not serviced optimally in terms of resources in static solution (yellowline) and dynamically optimized solution (green line).}
\label{fig:NumOfUE}
\centering
\end{figure*}

\subsection{ML Model Evaluation and Results}
In this subsection, we elaborate on the performance of the proposed intelligent scheme. Fig.~\ref{fig:NumOfUE} illustrates the performance of the proposed approach. In the first subgraph, the purple solid line represents the actual number of users in the slice with QCI1 and QCI9 indicators (QCI1 and QCI9 form a slice together), which varies over time given the dynamic traffic load of the network. In order to quantify the performance of the proposed approach, two scenarios are considered.

In the first scenario, we assume a constant limit on the number of users allowed in this slice (this is the approach currently implemented in real-world wireless networks). Correspondingly, the number of PRBs allocated to the slice is fixed and static over time. However, there might be instances where the actual number of users on the slice is lower or higher than the predefined fixed limit considered for resource orchestration. The second subgraph in Fig.~\ref{fig:NumOfUE}, demonstrates the number of under/over-served users in the slice, in this scenario. It is observed that there is a considerable amount of resources that are under-utilized on this slice while it can ideally be reserved for other slices having more resource needs.

On a separate scenario, we consider an adaptive limit on the number of users allowed in this slice (green line in the first subgraph). This adaptive limit is determined based on the traffic predicted (output of the RNN model) on the slice. Subsequently, the PRBs are re-orchestrated based on the adaptive limit on the slice for every time frame. It is notable that setting a high resolution time frame increases the amount of signaling and computation to perform the resource allocation. While considering a low resolution time frame might lead to inefficient allocation of the resources. Adjusting this trade-off is beyond the scope of this work. As observed in the third subgraph, the number of over/under-served uses has reduced significantly as resources are adjusted among slices in an adaptive manner. The last subgraph represents the total number of UEs that have been non-optimally served. It is worth noting that while reducing the under-served users are of importance to prevent SLA violation, decreasing the over-served users matter from network perspective in order to keep the expenditures low. A summation of under-served and over-served users is indicated as \emph{non-optimally served} UEs in Fig.~\ref{fig:NumOfUE}.
%\vspace{0.5cm}
%%%%%%%%%%%%%%%%
\section{Summary and Concluding Remarks}
\label{sec:Conclusion}
%%%%%%%%%%%%%%%%
This article provides an intelligent closed-loop SLA assurance scheme for O-RAN slicing. The O-RAN architecture enhances traditional network function with embedded intelligence and open and standard interfaces. Building on the two core principles of openness and intelligence, O-RAN facilitates agile deployment and fast optimization of the slices that cater for, potentially, different verticals.  We develop an intelligent and automated SLA assurance scheme through proactively re-orchestrating the resources over different slices. To this purpose, we utilize LSTM neural network to predict
the number of users, volume of  data  and  resource  utilization  on  each slice. The effectiveness of the predictor model was demonstrated using a real-world traffic dataset that is taken from a cellular network in India. However, feedback loop to the ML model considered is under investigation and presupposed to be our future work. Based upon the predicted volume of traffic and resource utilization of each slice, available resources are dynamically adjusted to handle an under-served slice. The proposed scheme is carried out in a closed-loop and automated fashion without human intervention. A high-level architecture and the corresponding flow are also provided.

Utilizing the proposed scheme on top of O-RAN architecture brings in operational  savings as a result of incorporating  intelligence  and  automation. Operators  can  mitigate a lot of problems such as congestion,  mobility management,  etc.  well ahead of time. Adoption of O-RAN gives an opportunity to the operators to integrate different intelligent solutions from different providers and create more competition. This  subsequently  brings  down  the capital and operational expenditures.
\bibliographystyle{IEEEtran}
\bibliography{IEEEabrv,Ref}
\vskip 0pt plus -1fil
\vspace{-0.25cm}
\begin{IEEEbiography}{Joseph Thaliath}
is currently working as Senior Chief Engineer at Samsung R\&D Bangalore, India.
%He received his M.Tech from IIIT, Bangalore in 2008, and his BE from Anna University, Chennai in 2006. His research interest includes O-RAN, vRAN, and Networks Slicing.
\end{IEEEbiography}
\vskip 0pt plus -1fil
\begin{IEEEbiography}{Solmaz Niknam} is currently a postdoctoral associate at Virginia Tech.
%received her B.Sc. degree (1st class honor) in Electrical Engineering from Shiraz University of Technology, Shiraz, Iran, in 2010, her M.Sc. degree in Electrical Engineering from Iran University of Science and Technology, Tehran, Iran, in 2012 and her Ph.D. degree from Kansas State University, KS, USA in 2018. During her Ph.D., she was a recipient of the Kansas Ph.D. students Fellowship. She is currently a postdoctoral associate at Virginia Tech. Her research interests include wireless communication with emphasis on heterogeneous networks, 5G mmWave networks and ML in communication.
%received her B.Sc. degree (1st class honor)  from Shiraz University of Technology, Shiraz, Iran, in 2010, her M.Sc. degree from Iran University of Science and Technology, Tehran, Iran, in 2012 and her Ph.D. degree from Kansas State University, KS, USA in 2018, all in Electrical Engineering. During her Ph.D., she was a recipient of the Kansas Ph.D. students Fellowship. After serving as a Postdoctoral Associate at Virginia Tech, she joined Qualcomm Technologies Inc. where she is currently a Senior Modem System Engineer.
%Her research interests include wireless communication with emphasis on 5G mm-wave networks and ML in communication.
\end{IEEEbiography}
\vskip 0pt plus -1fil
\vspace{-0.25cm}
\begin{IEEEbiography}{Sukhdeep Singh}
is currently working as Chief Engineer (Technical Manager) at Samsung R\&D Bangalore, India.
%He received his Ph.D. from Sungkyunkwan University, South Korea in 2016. His research interest includes 4G/5G RAN system design, O-RAN, vRAN, Cloud Native and TCP/QUIC for 5G/6G cellular networks.
\end{IEEEbiography}
\vskip 0pt plus -1fil
\vspace{-0.25cm}
\begin{IEEEbiography}{Rahul Banerji}
is currently working as Senior Software Engineer at Samsung R\&D India Bangalore.
%He received his B.E. from BITS Pilani, India in 2018. His research interest includes ML modeling and simulations for Next Generation Mobile Networks, Software Design of vRAN and O-RAN systems, Service Management and Operation for 5G wireless.
\end{IEEEbiography}
\vskip 0pt plus -1fil
\vspace{-0.25cm}
\begin{IEEEbiography}{Navrati Saxena}
is an assistant professor in the Department of Computer Science, San Jose State University, CA, USA.
%She completed her Ph.D. in the Department of Information and Telecommunication, University of Trento, Italy. Her prime research interests involve 4G/5G wireless, IoT, smart grids, and smart environments.
\end{IEEEbiography}
\vskip 0pt plus -1fil
\vspace{-0.25cm}
\begin{IEEEbiography}{Harpreet S. Dhillon} is an Associate Professor of Electrical and Computer Engineering and the Elizabeth and James E. Turner Jr. '56 Faculty Fellow at Virginia Tech.
%He received his B.Tech. degree from IIT Guwahati in 2008, his M.S. degree from Virginia Tech in 2010, and his Ph.D. degree from the University of Texas at Austin in 2013, all in Electrical Engineering. His research interests include communication theory, wireless networks, stochastic geometry, and machine learning. He is a Clarivate Analytics Highly Cited Researcher and a recipient of five best paper awards. He serves as an Editor for three IEEE journals.
\end{IEEEbiography}
\vskip 0pt plus -1fil
\vspace{-0.25cm}
\begin{IEEEbiography}{Jeffrey H. Reed} is currently the Willis G. Worcester Professor with the Bradley Department of Electrical and Computer Engineering, Virginia Tech.
%He received his B.S., M.S., and Ph.D. degrees from the University of California, Davis, CA, USA, in 1979, 1980 and 1987, respectively. He is the founder of Wireless@Virginia Tech and the Founding Faculty Member of the Ted and Karyn Hume Center for National Security and Technology. He is a Fellow of the IEEE for his contributions to software radio.
\vskip 0pt plus -1fil
\vspace{-0.25cm}
\begin{IEEEbiography}{Ali Kashif Bashir}
is an Associate Professor at the Department of Computing and Mathematics, Manchester Metropolitan University, United Kingdom.
%He is also with Visual Research Intelligent Center, University of Electronics Science and Technology of China (UESTC) as an Honorary Professor and Chief Adviser; with University of Science and Technology, Islamabad (NUST) as an Adjunct Professor, and with University of Guelph, Canada as Special Graduate Faculty. He is a senior member of IEEE, member of IEEE Industrial Electronic Society, member of ACM, and Distinguished Speaker of ACM. He received his Ph.D. in computer science and engineering from Korea University, South Korea. He has authored over 200 research articles; and received over 3 Million USD funding as PI and Co-PI from research bodies of South Korea, Japan, EU, UK and Middle East. His research interests include internet of things, wireless networks, distributed systems, network/cyber security, network function virtualization, machine learning, etc.
\end{IEEEbiography}
\vskip 0pt plus -1fil
\vspace{-0.25cm}
\begin{IEEEbiography}{Avinash Bhat}
is an Architect and Head of Global Pre-Sales (RAN) at Samsung R\&D India-Bangalore.
%He Joined Samsung in 2010. Prior to working in Samsung he was part of Motorola R\&D. He has an industry experience of 20 years. He completed his Bachelors (BE in ECE) from Manipal University in 2000. His current research interests include vRAN, O-RAN, ONAP and Network Slicing.
\end{IEEEbiography}
\vskip 0pt plus -1fil
\vspace{-0.25cm}
\begin{IEEEbiography}{Abhishek Roy}
is currently working as a senior technical manager at MediaTek.
%He received his Ph.D. degree in 2010 from Sungkyunkwan University, his M.S. degree in 2002 from the University of Texas at Arlington, and his B.E. degree in 2000 from Jadavpur University, India. He has strong professional skills in 4G/5G/6G wireless system design, New Radio unlicensed, IoT, cloud RAN, network modeling, and simulation.
\end{IEEEbiography}
\end{IEEEbiography}

\end{document}